\documentclass[preprintnumbers,superscriptaddress,showkeys,byrevtex]{revtex4}
\usepackage{amsmath,amsfonts,amssymb,amscd,amsxtra,amsthm}
\usepackage{graphicx}
\usepackage{epstopdf}
\usepackage{bm}
\usepackage{feynmp}
\usepackage{feynmp-auto}
\usepackage{amsmath,amsfonts,amssymb,amscd,amsxtra,amsthm}
\usepackage{orcidlink}
\usepackage{stackengine}
\begin{document}
\preprint{PKNU-NuHaTh-2024}
\title{Strangeness plus-one ($S=+1$) resonance-state $P^{+*}_0$ via $K^+n\to K^{*0}p$}
\author{Dayoung Lee\,\orcidlink{0009-0000-4828-3546}}
\email[E-mail: ]{ldyoung0421@pukyong.ac.kr}
\affiliation{Department of Physics, Pukyong National University (PKNU), Busan 48513, Korea}
\affiliation{Center for Extreme Nuclear Matters (CENuM), Korea University, Seoul 02841, Korea}
\author{Seung-il Nam$^*$\,\orcidlink{0000-0001-9603-9775}}
\email[E-mail: ]{sinam@pknu.ac.kr (corresponding author)}
\affiliation{Department of Physics, Pukyong National University (PKNU), Busan 48513, Korea}
\affiliation{Center for Extreme Nuclear Matters (CENuM), Korea University, Seoul 02841, Korea}
\affiliation{Asia Pacific Center for Theoretical Physics (APCTP), Pohang 37673, Korea}
\date{\today}
\begin{abstract}
In our current study, we delve into the peak-like structure observed during the reaction process of $K^+n\to K^{0}p$ at approximately $\sqrt{s}\sim2.5$ GeV. Our focus centers on exploring the potential $S=+1$ resonance $P^{+*}_0\equiv P^*_0$ as an excited state within the extended vector-meson and baryon ($VB$) antidecuplet. To achieve this aim, we employ the effective Lagrangian method in conjunction with the $(u,t)$-channel Regge approach, operating within the tree-level Born approximation. We thoroughly examine various spin-parity quantum numbers for the resonance, resulting in a compelling description of the data, where $M_{P^*_0}\approx2.5$ GeV and $\Gamma_{P^*_0}\approx100$ MeV. Furthermore, we propose an experimental technique to amplify the signal-to-noise ratio ($S/N$) for accurately measuring the resonance. Notably, our findings reveal that background interference diminishes significantly within the $K^*$ forward-scattering region in the center-of-mass frame when the $K^*$ is perpendicularly polarized to the reaction plane. Additionally, we explore the recoil-proton spin asymmetry to definitively determine the spin and parity of the resonance. This study stands to serve as a valuable reference for designing experimental setups aimed at investigating and comprehending exotic phenomena in QCD. Specifically, our insights will inform future J-PARC experiments, particularly those employing higher kaon beam energies.
\end{abstract}
\maketitle
\section{Introduction}
Quantum Chromodynamics (QCD) stands as the foundational principle governing the strong interactions among standard-model particles, portraying various hadrons through the non-perturbative interactions of quarks and gluons. It showcases confinement and asymptotic freedom within the framework of non-Abelian color-SU(3) gauge theory. While the minimal composition for baryons is represented by $qqq$ and for mesons by $q\bar{q}$, QCD does not forbid more intricate compositions such as $qqqq\bar{q}$ (pentaquark) and $q\bar{q}q\bar{q}$ (tetraquark), termed \textit{Exotics}. Experimental evidence supporting these exotics has accumulated over decades. The initial observation of the tetraquark meson $X(3872)$, which defies the simple quark model, dates back to 2003, reported by the Belle experiment~\cite{Belle:2003nnu}. Recently, its existence was reaffirmed by the LHCb~\cite{LHCb:2013kgk} and CMS~\cite{CMS:2021znk} experiments, specifying $J^{PS}=1^{++}$. Additionally, the Belle experiments reported evidence for $Y(4660)$ and $Z(4430)$ in 2007~\cite{Cotugno:2009ys}. Further tetraquark states have been identified via experiments conducted by LHCb, Belle, BES III, Fermilab, and others~\cite{Chen:2022asf}. Regarding baryons, LHCb reported the $P^+_c(4312,4440,4457)$, observed in decay into $J/\psi$ and $p$~\cite{LHCb:2015yax}. Recently, the same facility detected a novel pentaquark state with the strange-quark $s$ ($udsc\bar{c}$) in the decay of $B^-\to J/\psi\Lambda\bar{p}$~\cite{LHCb:2022ogu}. While not yet definitively confirmed, these pentaquarks can potentially be understood as bound states of vector-meson and baryon ($VB$), denoted as $P^+_c\sim D^*\Sigma_c$.

Remarkably, all experimentally observed and confirmed exotics exhibit heavy (charm) flavor. A theoretical rationale for this peculiarity concerning the stability of heavy pentaquarks was elucidated through color-spin interactions among quarks, contrasting with colorless hadronic interactions~\cite{Park:2018oib}. While the $S=+1$ light-pentaquark $\Theta^+\sim uudd\bar{s}$ was reported by the LEPS collaboration~\cite{LEPS:2008ghm}, supported by the non-topological chiral-quark soliton model~\cite{Diakonov:1997mm}, its existence remains disputed and unresolved due to varied outcomes across experimental facilities~\cite{PDG}. Furthermore, generating the $S=+1$ resonance dynamically proves challenging within the framework of Weinberg-Tomozawa (WT) type chiral interactions, primarily due to their repulsive nature. Nevertheless, the suggestion of an $S=0$ pentaquark-like bound-state $P^+_s\sim uuds\bar{s}\sim K^*\Sigma$, akin to $P^+_c$ in heavy flavor, was put forth by one of the present authors and collaborators~\cite{Nam:2021ayk,Khemchandani:2011et}, aiming to elucidate the bump structure observed near $\sqrt{s}\approx2.1$ GeV in $\gamma p\to\phi p$~\cite{Shim:2024myp}. In Ref.~\cite{Shim:2024myp}, the existence of $P^+_s$ was found to be crucial for reproducing the nontrivial structure observed in the spin-density matrix elements (SDMEs), in addition to successfully describing the angular-dependent cross-sections.

Given the current circumstances outlined above, it becomes of paramount importance to explore the potential for observing evidence of light-pentaquark states, either as bound states or resonances within the $VB$ framework, to further enhance our understanding of QCD regarding exotic particles. Additionally, even in the $S=+1$ channel, there may exist a resonance or bound-state pentaquark, despite the lack of support from chiral dynamics~\cite{KAN}. Hence, in the present study, we undertake a theoretical examination of relatively aged data pertaining to the $S=+1$ channel reaction process, such as $K^+n\to K^{*0}p$~\cite{Bologna-Glasgow-Rome-Trieste:1976hco}. Intriguingly, this reaction process exhibits a peak-like structure around $\sqrt{s}=2.5$ GeV, potentially indicative of an $S=+1$ resonance state, which we provisionally designate as $P^{+*}_0\equiv P^*_0$. We aim to elucidate this hypothetical pentaquark resonance $P^*_0$ in our current investigation. Drawing upon our previous work~\cite{Nam:2021ayk}, where we assigned the light-flavor $VB$ pentaquark as $P_s\sim qqqs\bar{s}$, assuming it as a member of the extended $VB$ antidecuplet flavor structure akin to $\Theta^+$, we can extend this assignment to other possible states: $P_{0}\sim qqqq\bar{s}$, $P_{1}\equiv P_s\sim qqqs\bar{s}$, $P_{2}\equiv P_{ss}\sim qqss\bar{s}$, and $P_{3}\equiv P_{sss}\sim qsss\bar{s}$, where $q$ represents the flavor-SU(2) light quarks ($u$ and $d$), being categorized by the numbers of the strange quarks they include. 

In our previous study~\cite{Nam:2021ayk}, we set the mass of $P_s$ to be approximately $2.071$ GeV, guided by the chiral-unitary approach~\cite{Khemchandani:2011et}. Assuming $P_s(2071)$ as the $S=0$ ground-state pentaquark of the extended $VB$ antidecuplet and considering the mass difference between antidecuplets due to the inertia moment of the non-topological chiral soliton, approximately $\sim180$ MeV~\cite{Diakonov:1997mm}, the ground-state $P_0$ mass is estimated to be around $1.890$ GeV. Consequently, we regard the hypothetical resonance in the vicinity of the bump structure at $2.5$ GeV as a resonance state of $P_0$, denoted as $P^*_0(2500)$. Furthermore, we highlight several advantageous features of the chosen process: Firstly, since all hadrons in the final state of $K^+n\to K^{*0}p\to \pi^{\mp} K^{\pm} p$ can be measured experimentally in principle, the Fermi-motion effects of the deuteron-target experiment can be minimized. Secondly, in the $s$ channel, it is possible to measure the resonance simultaneously coupling to $PB$ and $VB$ channels, allowing for the study of its production mechanism in two distinct dynamics. Lastly, this process involves the vector meson with polarizations, enabling the selection of specific quantum numbers for the resonance and control over its production mechanism.

To achieve this objective, we employ a straightforward tree-level Born approximation calculation utilizing the effective Lagrangian method. We account for the Regge trajectories in the $t$ and $u$ channels, incorporating phenomenological form factors to characterize the spatial extension of the hadrons. To replicate the observed peak-like structure in the total cross-section, potentially indicative of the $S=+1$ $VB$ resonance ($\equiv P^{+*}_0$), we introduce a relativistic Breit-Wigner type contribution with $M_R=2.5$ GeV and $\Gamma_R=100$ MeV. Our numerical calculations reveal that, alongside the resonance contribution in the $s$ channel, Reggeized $\pi$-exchange in the $t$ channel and $\Lambda(1116)$ one in the $u$ channel dominate the reaction process. We explore various spin-parity quantum numbers theoretically ($J^P=1/2^\pm,3/2^\pm$) since experimental data to determine them precisely are lacking. Notably, all tested spin-parity quantum numbers qualitatively reproduce the data well. To mitigate the signal-to-noise ($S/N$) ratio, we propose a specific combination of scattering angle and polarization ($\theta_{K^*}\sim0$ and $k_K\cdot\epsilon_{K^*}\sim0$) for future experiments, surpassing current programs like the K1.8 beam-line at J-PARC, which require a kaon beam energy of approximately $3.0$ GeV to access the $P^*_0$ region. Additionally, we recommend utilizing the recoil-proton spin asymmetry to ascertain the spin parity of the resonance.

The current study is structured as follows: Section II presents the theoretical framework. In Section III, we present the numerical results along with detailed explanations and discussions. Finally, the concluding section provides a summary and outlines future directions.
\section{Theoretical Framework}
\begin{figure}[t]
\includegraphics[width=16cm]{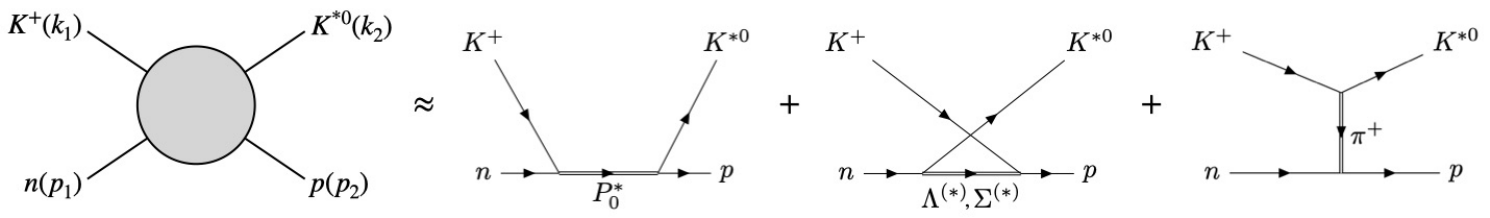}
\caption{Feynman diagrams pertinent to the process $K^+n\to K^{*0}p$ within the tree-level Born approximation are presented.}       
\label{FIG0}
\end{figure}

In this section, we will provide a concise overview of the theoretical framework governing the present reaction process. Relevant Feynman diagrams for the process are illustrated in Fig.~\ref{FIG0}, where the four momenta of the particles involved are also defined. The Yukawa interaction vertices are described as follows:
\begin{eqnarray}
\label{eq:EFLS}
\mathcal L_{PBB'}&=&\frac{g_{PBB'}}{M_K}\bar B'^{\mu}\Theta_{\mu\nu}(X)(\partial^{\nu}K)\Gamma_5B+\rm h.c.,
\cr
\mathcal L_{VBB'}&=&-\frac{ig_{VBB'}}{M_{K^*}}\bar B'^{\mu}\gamma^{\nu}(\partial_{\mu}K^*_{\nu}-\partial_{\nu}K^*_{\mu})\Gamma_5\gamma_5B+\rm h.c.,
\cr
\mathcal L_{PBB}&=&ig_{PBB}\bar{B}\Gamma_5\gamma_5K^{\dagger}B+{\rm h.c.},
\cr
\mathcal L_{VBB}&=&g_{VBB}\bar{B}\gamma^{\mu}\Gamma_5V^{\dagger}_{\mu}B+{\rm h.c.},
\cr
\mathcal L_{PPV}&=&ig_{PPV}V^{\mu}[P\partial_{\mu}P^{\dagger}-P^{\dagger}\partial_{\mu}P]+\rm h.c.,
\cr
\mathcal L_{PBB}&=&\Big(\frac{f_{PBB}}{M_P}\Big)\bar{B}{\gamma_5}(\rlap{/}{\partial}P)B.
\end{eqnarray}
In this context, $B'$, $B$, $P$, and $V$ denote the fields corresponding to spin-$3/2$, spin-$1/2$ baryons, pseudoscalar, and vector mesons, respectively. We utilize the notation $\Gamma_5=(I_{4\times4},\gamma_5)$ to represent the parity ($+,-$) of $B'$. The term $\Theta_{\mu\nu}(X)$ signifies the off-shell parameter $X$-dependent term within the Rarita-Schwinger formalism~\cite{Rarita:1941mf}.
\begin{eqnarray}
\label{eq:OFF}
\Theta_{\mu\nu}(X)=g_{\mu\nu}+X\gamma_{\mu}\gamma_{\nu}.
\end{eqnarray}
As illustrated in Fig.~\ref{FIG0}, our analysis encompasses the $P^*_0$ diagram in the $s$ channel, hyperon ($\Lambda^{(*)}$, $\Sigma^{(*)}$) exchange in the $u$ channel, and $\pi$ exchange in the $t$ channel. Given the absence of experimentally confirmed $S=+1$ baryon states to date, we exclusively focus on the hypothetical $VB$ pentaquark $P^*_0$. Concerning the $t$ channel, while there could be additional meson exchanges, such as the axial-vector meson $b_1(1235)$, we exclude them due to the scant information available regarding their decays~\cite{PDG}.

Following straightforward calculations, the invariant amplitudes for the Feynman diagrams depicted in Fig.~\ref{FIG0} are presented as follows:
\begin{eqnarray}
\label{eq:AMPS}
i\mathcal M_{{P^*_0}(s=1/2)}^s&=&-g_{K^*N{P^*_0}}g_{KN{P^*_0}}\bar{u}(p_2)\Gamma_5
\rlap{/}{\epsilon}\frac{\rlap{/}{q}_s+M_{P^*_0}}{s-M_{P^*_0}^2-iM_{P^*_0}\Gamma_{P^*_0}}\gamma_5\Gamma_5u(p_1),
\cr
i\mathcal M_{{P^*_0}(s=3/2)}^s&=&-\frac{ig_{K^*N{P^*_0}}g_{KN{P^*_0}}}{M_{K^*}M_K}\bar u(p_2)
\gamma_5\Gamma_5 (k_{2\mu}\epsilon_\nu-k_{2\nu}\epsilon_\mu)\gamma^{\nu} 
\frac{\left(\rlap{/}{q}_s+M_{P^*_0}\right)\bar{g}^{\mu\alpha}(q_s)}{s-M_{P^*_0}^2+iM_{P^*_0}\Gamma_{P^*_0}} 
\big(g_{\alpha\beta}+X\gamma_{\alpha}\gamma_{\beta}\big) k_1^\beta \Gamma_5u(p_1),
\cr
i\mathcal M^u_{Y}&=&-\frac{i}{2}g_{KNY}g_{K^*NY}\bar u(p_2)\Gamma_5\frac{(\rlap{/}{q}_u+M_Y)\rlap{/}{\epsilon}}{u-M_Y^2-iM_Y\Gamma_Y} \Gamma_5\gamma_5u(p_1),
\cr
i\mathcal M_{\Sigma^*(1385)}^u&=&\frac{i}{2} \frac{g_{K^*N\Sigma^*}}{M_{K^*}} \frac{g_{KN\Sigma^*}}{M_K} \bar u(p_2) k_1^{\nu}\left(g_{\mu\nu}+X\gamma_{\mu}\gamma_{\nu}\right) \frac{\bar{g}^{\mu\alpha}(q_u)\left(\rlap{/}{q}_u+M_{\Sigma^*}\right)}{u-M_{\Sigma^*}^2+iM_{\Sigma^*}\Gamma_{\Sigma^*}} 
		\gamma^{\beta} \left(k_{2\alpha}\epsilon_{\beta}-k_{2\beta}\epsilon_{\alpha}\right) \gamma_5 u(p_1),
\cr
i\mathcal M_{\pi}^t&=&-\frac{2if_{\pi NN}\ g_{PPV}}{M_{\pi}}\overline u(p_2){\gamma_5}
		\frac{\rlap{/}{q}_t(\epsilon\cdot k_1)}{t-M_\pi^2}u(p_1).
\end{eqnarray}
Here, the notation $(s,u,t)=(q^2_s,q^2_u,q^2_t)$ denotes the Mandelstam variables. In our calculations, we utilize the spin-$1/2$ hyperons $Y=\Lambda(1116,1/2^+)$, $\Sigma(1192,1/2^+)$, $\Lambda(1405,1/2^-)$, and $\Lambda(1670,1/2^-)$, along with the hyperon resonance $\Sigma^*(1385,3/2^+)$. The polarization vector of $K^*$ is represented by $\epsilon_\mu$. The spin sum of the Rarita-Schwinger field yields the following expression:
\begin{eqnarray}
\label{eq:SPINSUM}
\bar{g}^{\mu\nu}(q)=g^{\mu\nu}-\frac{1}{3}\gamma^{\mu}\gamma^{\nu}-\frac{2}{3M^2}q^{\mu}q^{\nu}+\frac{q^{\mu}\gamma^{\nu}+q^{\nu}\gamma^{\mu}}{3M}.
\end{eqnarray}
Here, $M$ and $\Gamma$ represent the mass and full decay width of the particle corresponding to the field. All relevant couplings are established based on experimental data and theoretical frameworks~\cite{PDG,Khemchandani:2011et,Stoks:1994nh}, as summarized in Table~\ref{TAB1}, except for those for $P^*_0$. To streamline our analysis, we denote the combined coupling as $g_{P^*0}\equiv g_{K^*N{P^*_0}}g_{KN{P^*_0}}$ hereafter.
\begin{table}[b]
\begin{tabular}{p{2.5cm}|p{2.5cm}|p{2.5cm}|p{2.5cm}}
\hfil Hyperon $(Y)$& \hfil$g_{K^*NY}$ &\hfil $g_{KNY}$ & \hfil Combined \\ 
\hline
\hfil$\Lambda(1116,1/2^+)$ &\hfil $-4.26$ & \hfil$-13.4$ & \hfil$57.08$  \\
\hfil$\Lambda^*(1405,1/2^-)$ & \hfil$0.2-2.7i$ & \hfil$2.5+0.9i$ & \hfil$2.93-6.57i$ \\
\hfil$\Lambda^*(1670,1/2^-)$ & \hfil$-0.2+0.8i$ & \hfil$0.2-0.6i$ & \hfil$-0.52-0.04i$ \\ 
\hfil$\Sigma(1192,1/2^+)$ &\hfil $-2.46$ & \hfil$4.09$ & \hfil$-10.06$ \\ 
\hfil$\Sigma^*(1385,3/2^+)$ &\hfil $-5.48$ & \hfil$-6.94$ & \hfil$38.03$\\
\hline
\hline
\hfil $\pi$ &\hfil  $f_{\pi NN}=0.989$ & \hfil $g_{\pi K K^*}=3.76$ & \hfil $3.72$ 
\end{tabular}
\caption{Strong couplings for the hyperon and pion vertices and their combinations derived from theoretical models and experimental data~\cite{PDG,Khemchandani:2011et, Stoks:1994nh}. }
\label{TAB1}
\end{table}

To account for the spatial extension of the hadrons, we incorporate a phenomenological form factor, defined as follows:
\begin{eqnarray}
\label{eq:FF}
F(x)=\frac{\Lambda^4}{\Lambda^4+(x-M^2_x)}.
\end{eqnarray}
In this expression, $x$ represents the Mandelstam variable, and $M_x$ denotes the off-shell mass of the corresponding particle. The cutoff parameter $\Lambda$ will be adjusted to accurately reproduce the experimental data.

Given that the available data from Ref.~\cite{Bologna-Glasgow-Rome-Trieste:1976hco} extends up to $\sqrt{s}\sim5$ GeV, which approaches the asymptotic limit $s\to\infty$, it becomes imperative to account for the higher-spin states of the exchange particles. To address this, we incorporate the $t$- and $u$-channel Regge trajectories. Following the methodology outlined in Ref.~\cite{Kim:2023jij}, the $u$-channel amplitudes for the $\Lambda(1116)$, $\Sigma(1192)$, and $\Sigma^*(1385)$ are adjusted using the Regge approach, wherein each trajectory $\alpha(x)$ is characterized by the intercept $\alpha'$. 
\begin{eqnarray}
\label{eq:REGGE}
\tilde{\mathcal M}_{\Lambda,\Sigma}^u(s,u)&=&C_{\Lambda,\Sigma}(u)\mathcal M'^u_{\Lambda,\Sigma}\left(\frac{s}{s_{\Lambda,\Sigma}}\right)^{\alpha_{\Lambda,\Sigma}(u)-\frac{1}{2}}\Gamma\left[\frac{1}{2}-\alpha_{\Lambda,\Sigma}(u)\right]\alpha'_\Lambda,
\cr
\tilde{\mathcal M}_{\Sigma^*}^u(s,u)&=&C_{\Sigma^*}(u)\mathcal M'^{u}_{\Sigma^*}\left(\frac{s}{s_{\Sigma^*}}\right)^{\alpha_{\Sigma^*}(u)-\frac{3}{2}}\Gamma\left[\frac{3}{2}-\alpha_{\Sigma^*}(u)\right]\alpha'_{\Sigma^*}.
\end{eqnarray}
Note that the amplitude $\mathcal{M}'$ represents the expression in Eq.~(\ref{eq:AMPS}) after removing the denominator part of the propagator. The trajectories are specified as $\alpha_{\Lambda}(u)=-0.65+0.94u$, $\alpha_{\Sigma}(u)=-0.79+0.87u$, and $\alpha_{\Sigma^*}(u)=-0.27+0.9u$. Additionally, the phenomenological momentum-dependent hadronic coefficient is defined as follows:
\begin{eqnarray}
\label{eq:REGGE1}
C_{Y}(u)=\left[\frac{\epsilon_{Y}\Lambda_{Y}^2}{\Lambda_{Y}^2-u}\right]^2.
\end{eqnarray}
Finally, the $t$-channel Reggeized amplitude for the $\pi$ reads
\begin{eqnarray}
\label{eq:REGGE2}
\tilde{\mathcal M}_{\pi}^t(s,u)&=&C_\pi(u)\mathcal M'^t_\pi\left(\frac{s}{s_\pi}\right)^{\alpha_\pi(t)}
\frac{\pi\alpha'_\pi}{\Gamma[\alpha_X(t)+1]\sin[\pi\alpha_\pi(t)]}.
\end{eqnarray}
Here, $C_\pi(t)=a_\pi/(1-t/\Lambda_\pi^2)^2$. The Regge parameters and cutoffs for the form factors are provided in Table~\ref{TAB2}. It is important to mention that we adopt $X=0$ for simplicity in our analysis.

\begin{table}[b]
\begin{tabular}{p{2.5cm}|p{2.5cm}|p{2.5cm}|p{2.5cm}}
\hfil Hyperon ($Y$)& 
\hfil$\eta_{Y}$ & \hfil$s_{Y}\,[\rm GeV^2]$ & \hfil$\Lambda_{Y}\,[\rm GeV]$  \\ 
\hline 
\hfil$\Lambda(1116,1/2^+)$ & \hfil$2.60$ & \hfil$1.10$ & \hfil$0.45$   \\
\hfil$\Sigma(1192,1/2^+)$ & \hfil$0.66$ & \hfil$1.10$ & \hfil$0.45$  \\ 
\hfil$\Sigma^*(1385,3/2^+)$ & \hfil$0.66$ & \hfil$1.10$ & \hfil$0.45$ \\
\hline
\hline
\hfil$\pi$ & \hfil$\alpha_{\pi}=0.53$ & \hfil$s_\pi=1.00$ & \hfil$\Lambda_\pi=1.00$
\end{tabular}
\caption{Parameters for the Regge approach and cutoffs in the present work~\cite{Kim:2023jij}.}
\label{TAB2}
\end{table}
\section{Numerical results and Discussions}
\begin{figure}[t]
\begin{tabular}{cc}
\topinset{(a)}{\includegraphics[width= 8.5cm]{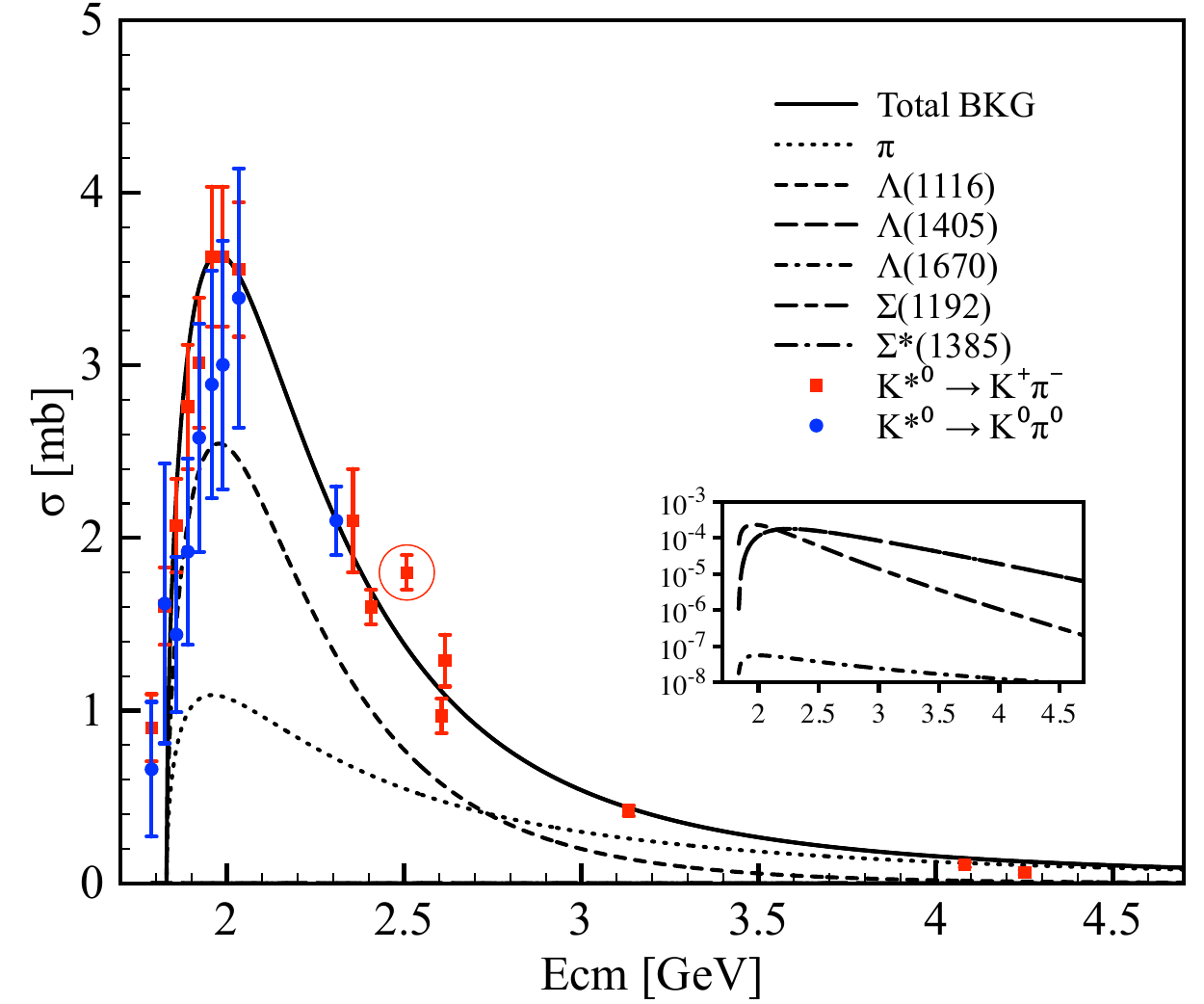}}{0.5cm}{-2cm}
\topinset{(b)}{\includegraphics[width= 8.5cm]{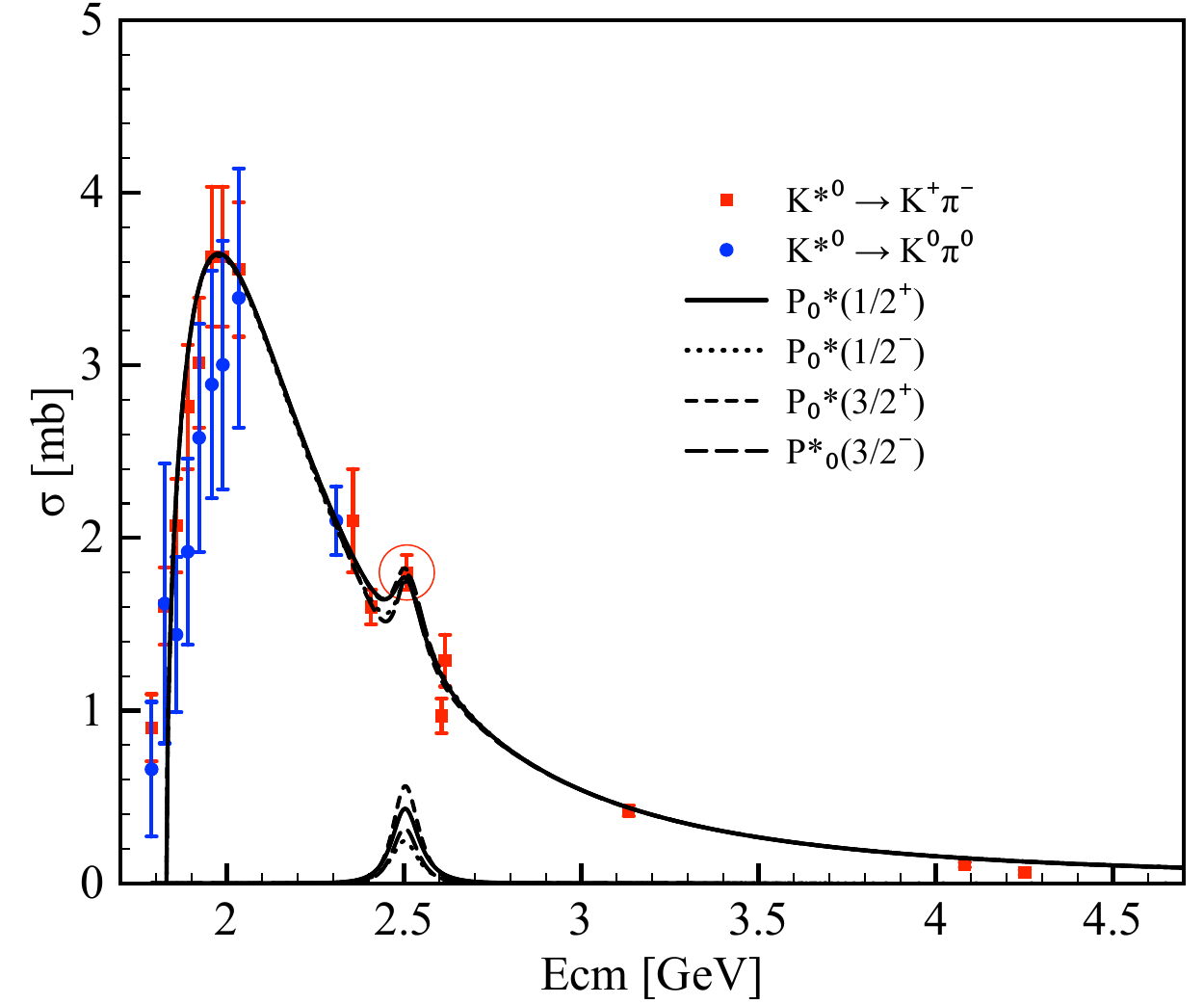}}{0.5cm}{-2cm}
\end{tabular}
\caption{(a) Total cross-sections for the non-resonant contributions as functions of $E_\mathrm{cm}$, serving as the background (BKG), including the total (solid), $\pi$ (dotted), $\Lambda(1116)$ (dashed), $\Lambda(1670)$ (long-dashed), $\Lambda(1670)$ (dot-dashed), $\Sigma(1192)$ (long-short-dashed), and $\Sigma^*(1385)$ (long-dot-dashed) components. (b) Total cross-sections with four different spin-parity quantum numbers of the $P^*_0$: $J^P=1/2^+$ (solid), $J^P=1/2^-$ (dotted), $J^P=3/2^+$ (dashed), and $J^P=3/2^-$ (long-dashed). Experimental data are obtained from various sources, including Refs.~\cite{Bologna-Glasgow-Rome-Trieste:1976hco,Baker:1975,Goldhaber:1965,Buchner:1972,Bassompierre:1970,Hendrickx:1976,Cords:1971}. Data points that deviate from the background description are marked with hollow circles.}       
\label{FIG1}
\end{figure}

In this section, we present the numerical results for the reaction process along with detailed discussions. The cutoffs for the form factors in Eq.~(\ref{eq:FF}) are chosen to be $1.0$ GeV and $0.45$ GeV for the $(s,t)$ and $u$ channels, respectively, by fitting the data as shown in Table~\ref{TAB2}. Firstly, in panel (a) of Fig.~\ref{FIG1}, we depict the total cross-sections as functions of the center-of-mass energy (cm)($E_\mathrm{cm}$), showcasing the non-resonant $u$- and $t$-channel contributions separately. The lines represent contributions from the total (solid), $\pi$ (dotted), $\Lambda(1116)$ (dashed), $\Lambda(1670)$ (long-dashed), $\Lambda(1670)$ (dot-dashed), $\Sigma(1192)$ (long-short-dashed), and $\Sigma^*(1385)$ (long-dot-dashed).  Experimental data are obtained from both charged (circle) and neutral (square) channels~\cite{Bologna-Glasgow-Rome-Trieste:1976hco,Baker:1975,Goldhaber:1965,Buchner:1972,Bassompierre:1970,Hendrickx:1976,Cords:1971}. We observe that the $\Lambda(1116)$ exchange in the $u$-channel is crucial for reproducing the strength of the cross-section near the threshold, while the contributions from other hyperons are almost negligible, as inferred from their coupling strengths with the Regge coefficients, as depicted in Tables~\ref{TAB1} and \ref{TAB2}. Conversely, the $\pi$-exchange contribution in the $t$ channel is significant for describing the relatively higher-energy region. However, these non-resonant contributions fail to explain the peak-like structure at $E_\mathrm{cm}\approx2.5$ GeV. In the figure, data points beyond the description provided by the background (BKG) are denoted with a hollow circle. Several explanations can potentially account for the observed peak-like structure in the cross-sections: 1) Single or combinations of resonances, which are the focus of our current investigation; 2) Opening of meson-baryon scattering channels~\cite{Ahn:2019rdr}; 3) Interferences of scattering amplitudes~\cite{LEPS:2016ljn}; and 4) Non-trivial kinematic singularities~\cite{Szczepaniak:2015eza}.

Considering the scattering of ground-state vector-meson and baryon multiplets ($8,10,\bar{10}$) in the $S=+1$ channel, the possibility of channel opening appears less likely due to their masses, which are approximately $M_\mathrm{max}=M_{K^*}+M_\Delta\approx2.12$ GeV. Although scatterings involving higher multiplets may contribute, this possibility is beyond the scope of our current investigation. We also explore the interferences between the $u$- and $t$-channel amplitudes with various phase factors, yielding no significant structures. Another potential source of interference could arise from crossing the invariant-mass bands on the Dalitz plot for $K^+n\to \pi^+K^-p$, as a function of $M_{\pi^+K^-}$ and $M_{K^-p}$, for instance. It's worth noting that similar considerations regarding interference possibilities have been explored theoretically and experimentally for $\gamma p\to\pi^+ K^-p$, with a focus on $\phi$ production~\cite{LEPS:2016ljn}. It was found that interference effects were negligible. Nontrivial effects, such as the triangle singularity~\cite{Szczepaniak:2015eza}, could also be introduced. For example, one might consider a triangle diagram consisting of the $P$-$V$-$B$ internal lines in the present work, where higher-mass meson-baryon cuts can lead to singularities. However, achieving a peak at $E_\mathrm{cm}\approx2.5$ GeV would require very high-mass hadrons. Nonetheless, this remains an intriguing subject for future exploration, and we leave it for subsequent studies.

Therefore, our focus lies in identifying the presence of a resonance within the observed structure. It's important to note that we explore four different spin-parity scenarios for $P^*_0$, namely $J^P=1/2^\pm,\,3/2^\pm$, as there is currently no experimental data available for reference. It has been verified that all these scenarios can qualitatively reproduce the peak-like structure effectively by adjusting the combined couplings as follows:
\begin{equation}
g^{1/2^+}_{P^*_0}=0.65,\,\,\,\,	
g^{1/2^-}_{P^*_0}=0.15,\,\,\,\,	
g^{3/2^+}_{P^*_0}=0.70,\,\,\,\,	
g^{3/2^-}_{P^*_0}=0.23.
\end{equation}
It's worth noting that the full decay width of $P^*_0$ remains consistent across all spin-parity quantum numbers, set at $\Gamma_{P^*_0}=100$ MeV. In panel (b) of Fig.~\ref{FIG1}, we present the total cross-sections for each case separately: $J^P=1/2^+$ (solid line), $J^P=1/2^-$ (dotted line), $J^P=3/2^+$ (dashed line), and $J^P=3/2^-$ (long-dashed line).

\begin{figure}[t]
\begin{tabular}{cc}
\topinset{(a)}{\includegraphics[width= 8.5cm]{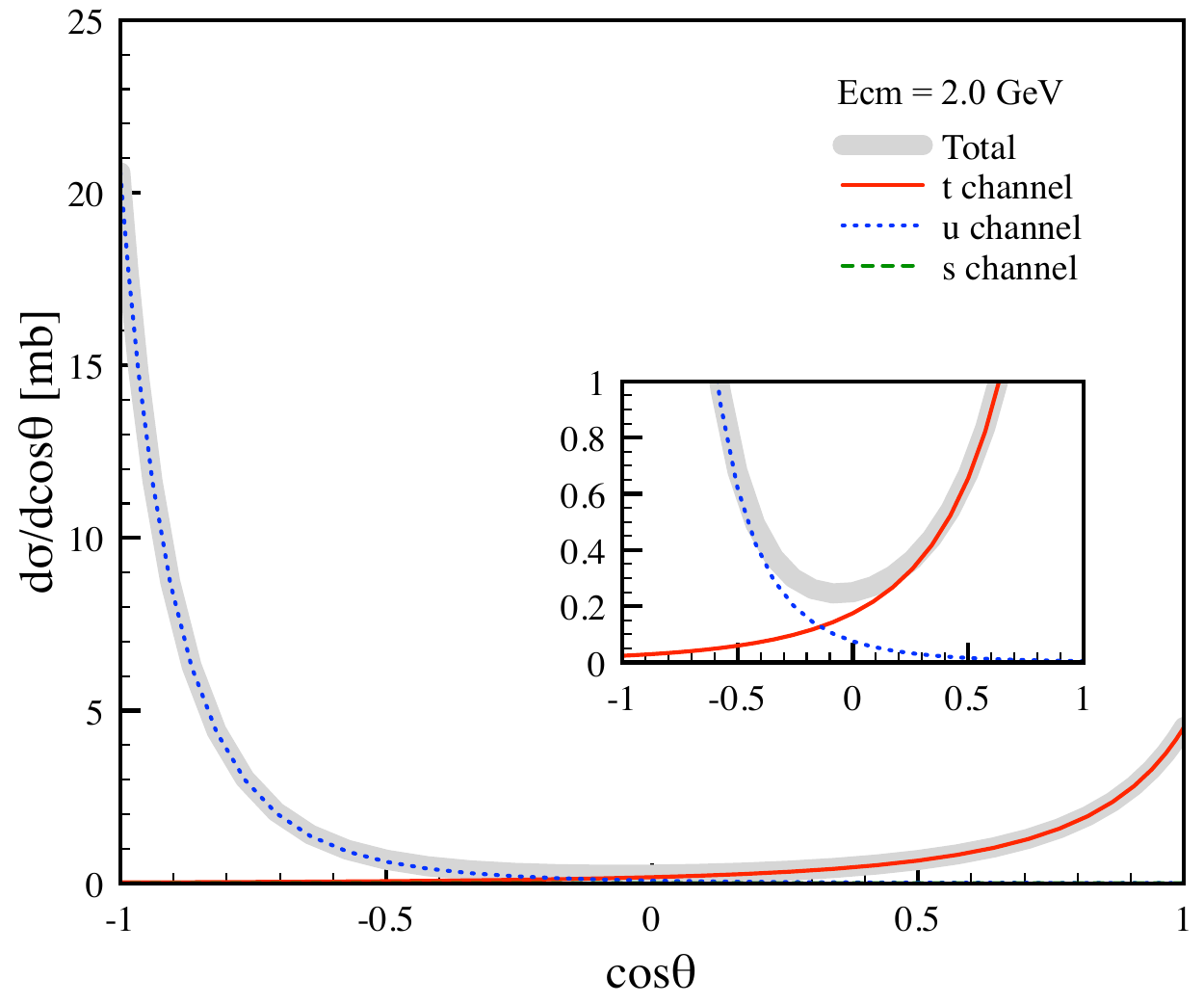}}{0.5cm}{-2cm}
\topinset{(b)}{\includegraphics[width= 8.5cm]{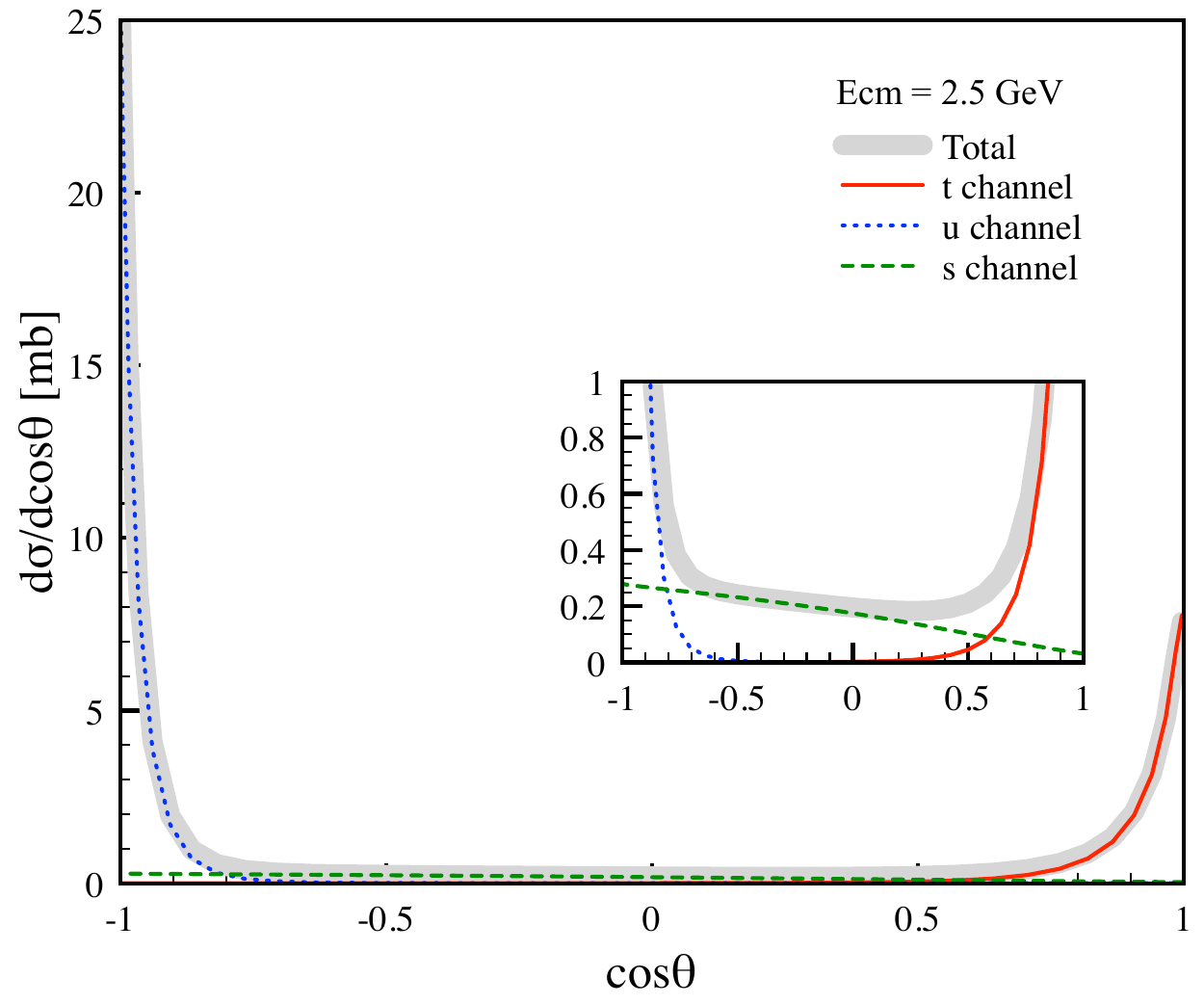}}{0.5cm}{-2cm}
\end{tabular}
\begin{tabular}{cc}
\topinset{(c)}{\includegraphics[width= 8.5cm]{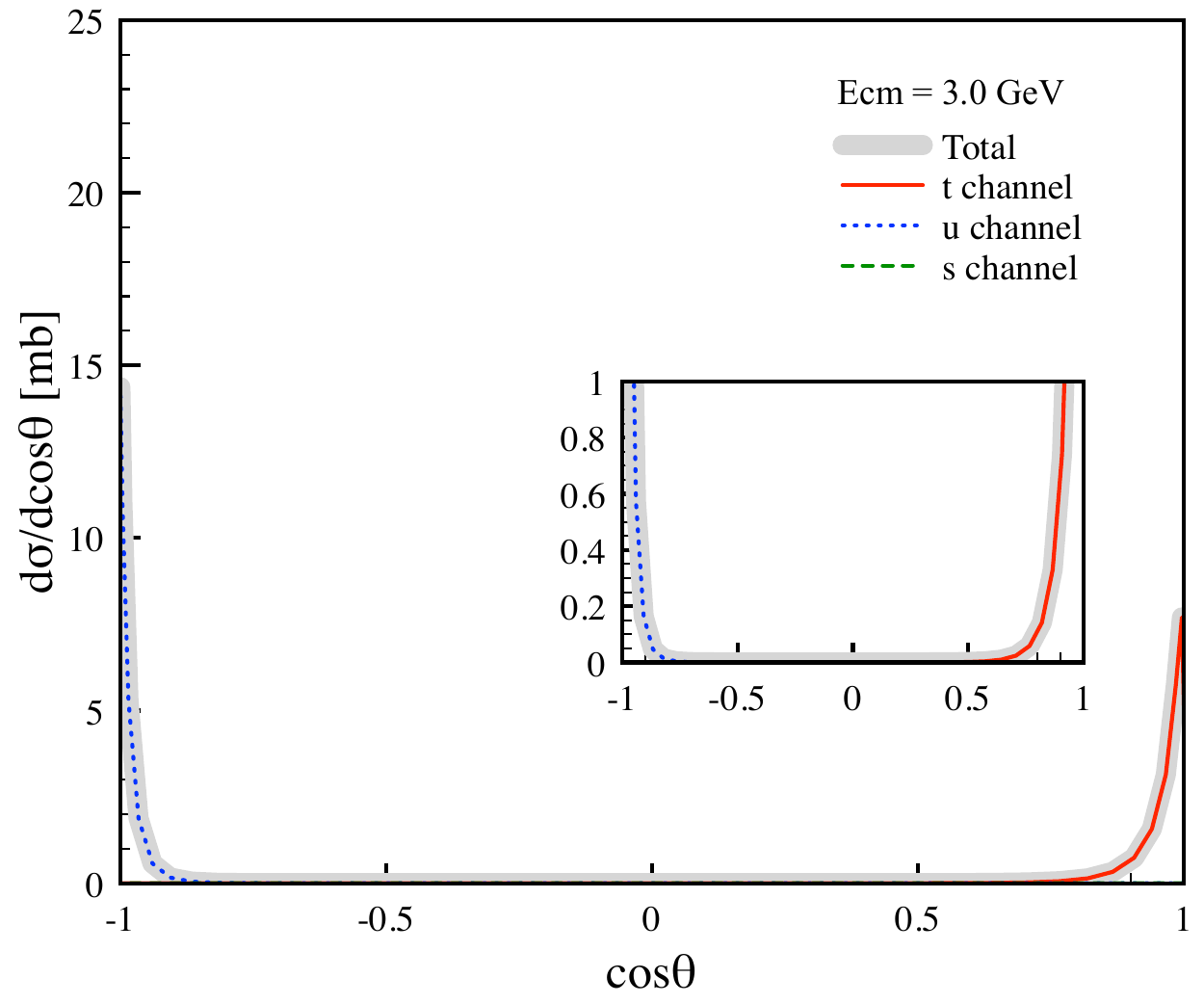}}{0.5cm}{-2cm}
\topinset{(d)}{\includegraphics[width= 9cm]{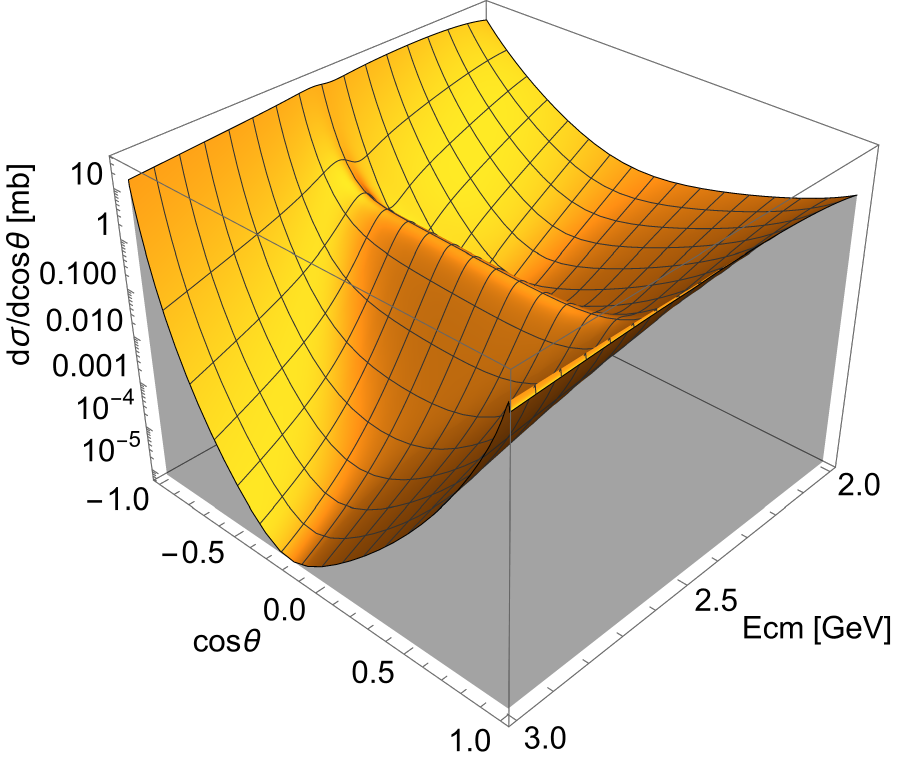}}{0.5cm}{-2cm}
\end{tabular}
\caption{The differential cross-sections for the angular dependence $d\sigma/d\cos\theta$ [mb] at $E_\mathrm{cm}=(2.0,2.5,3.0)$ GeV are depicted in panels (a), (b), and (c) as functions of $\cos\theta$, illustrating each contribution. Here, $\theta$ represents the scattering angle of the final state $K^*$ in the cm frame. Panel (d) displays $d\sigma/d\cos\theta$ as a function of $E_\mathrm{cm}$ and $\cos\theta$. In this panel, we exclusively showcase the results for $P^*_0(3/2^-)$, as other quantum-number states exhibit negligible differences.}       
\label{FIG2}
\end{figure}

In Fig.~\ref{FIG2}, the numerical results for the differential cross-sections for the angular dependence $d\sigma/d\cos\theta$ are presented for $E_\mathrm{cm}=(2.0,2.5,3.0)$ GeV in panels (a), (b), and (c), respectively, illustrating each contribution. Here, $\theta$ denotes the scattering angle of the final state $K^*$ in the cm frame. Notably, strong forward and backward scatterings are observed from the hyperons in the $u$ channel and from $\pi$ in the $t$ channel, respectively. Conversely, the $s$-channel contribution yields almost flat curves, as anticipated. Around $E_\mathrm{cm}=2.5$ GeV, corresponding to the resonance region (where we use $3/2^-$), a slight yet discernible angular dependence emerges. As the energy surpasses the resonance region, the angular dependence markedly separates into forward- and backward-scattering regions. Panel (d) presents the angular dependence as a function of $E_\mathrm{cm}$ and $\cos\theta$. Notably, the resonance signal becomes more pronounced around $\cos\theta=0$, where the contributions from the $t$ and $u$ channels are diminished or negligible. These observations suggest a potential method for experimentally isolating the resonance $P^*_0$ contribution, which we will discuss below.

\begin{figure}[t]
\begin{tabular}{cc}
\topinset{(a)}{\includegraphics[width= 8.5cm]{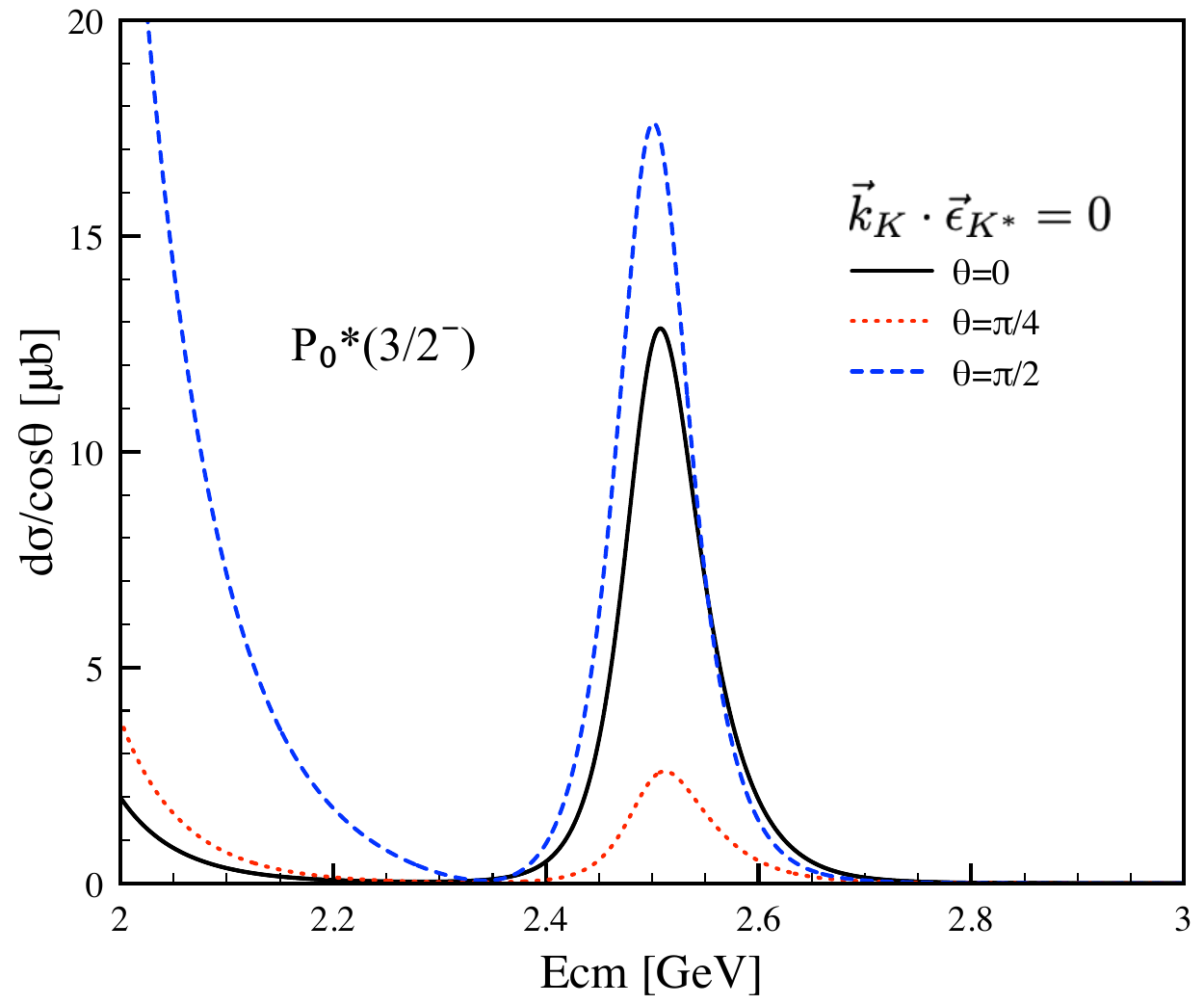}}{0.5cm}{-2cm}
\topinset{(b)}{\includegraphics[width= 9cm]{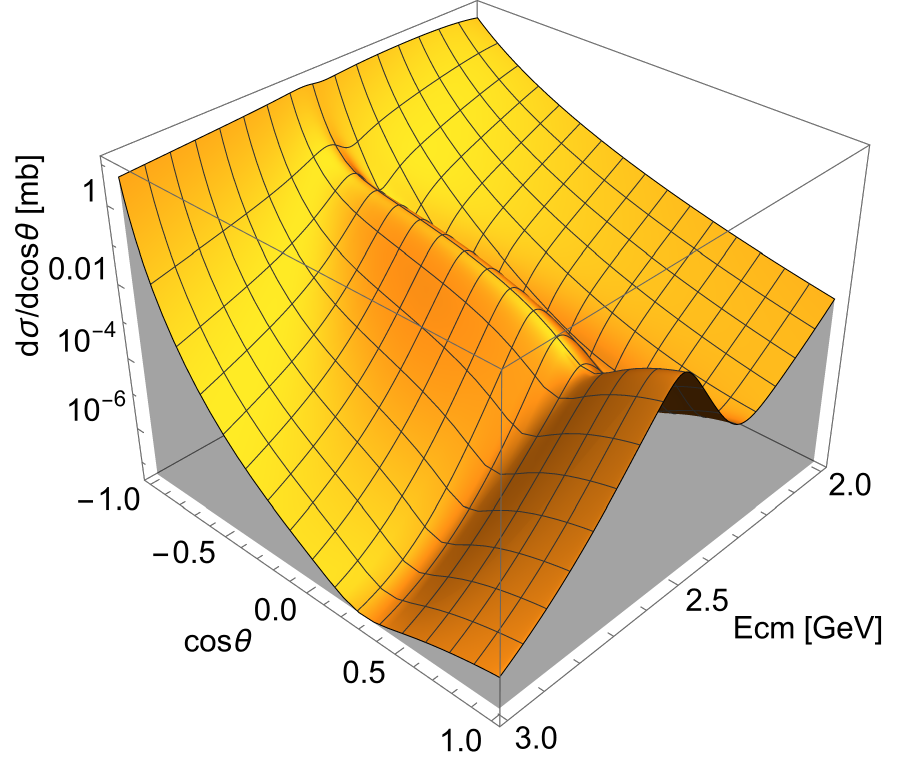}}{1.5cm}{3cm}
\end{tabular}
\caption{(a) Differential cross-sections $d\sigma/d\cos\theta$ for $\bm{k}_K\cdot\bm{\epsilon}_{K^*}=0$ for $\theta=0$ (solid), $\pi/4$ (dotted), and $\pi/2$ (dashed) in the cm frame. Here, we only present the results for the $P^*_0(3/2^-)$, as other quantum-number states do not significantly differ. (b) The same as (a), plotted as a function of $\cos\theta$ and $E_\mathrm{cm}$ for $\bm{k}K\cdot\bm{\epsilon}{K^*}=0$ at $\theta=0$.}       
\label{FIG3}
\end{figure}

Now, we are in a position to discuss how to enhance the $S/N$ for the $P^*_0$ resonance. As previously illustrated, the dominant contributions in the present reaction process originate from the $u$- and $t$-channel interactions. Therefore, by focusing solely on the forward-scattering region, where the hyperon-induced backward scattering is negligible, we can effectively eliminate this background. The subsequent step involves reducing the contribution from the $t$-channel $\pi$-exchange. Remarkably, due to its vector-meson nature, the invariant amplitude for the $\pi$ exchange in Eq.~(\ref{eq:AMPS}) is directly proportional to $\bm{k}_K\cdot\bm{\epsilon}_{K^*}$, where $\bm{k}_K$ represents the three-momentum of the initial $K^+$ meson. Consequently, when the polarization of $K^*$ is fixed perpendicular to the reaction plane, where all particle momenta are defined, the $\pi$-exchange contribution diminishes. In panel (a) of Fig.~\ref{FIG3}, we display the differential cross-sections $d\sigma/d\cos\theta$ as a function of $E_\mathrm{cm}$ for $\bm{k}_K\cdot\bm{\epsilon}_{K^*}=0$ at $\theta=0$ (solid), $\pi/4$ (dotted), and $\pi/2$ (dashed) in the cm frame. Here, we specifically focus on the result for the $P^*_0(3/2^-)$, as other quantum-number states exhibit minimal variation. At $\theta=0$, where $K^*$ is scattered forwardly, the $P^*_0$ signal is isolated with minimal background contamination. As the angle increases, the $S/N$ value worsens, particularly at $\theta=\pi/4$. Further increasing it to $\theta=\pi/2$ enhances the peak due to interference between the resonance and the $u$-channel contribution, although the background contamination also increases. In panel (b) of Fig.~\ref{FIG3}, we plot the differential cross-section as a function of $\cos\theta$ and $E_\mathrm{cm}$. Across a wide range of angles ($-0.5\lesssim\cos\theta\le1$), clear resonance signals are observed, facilitating their measurement in experiments. Additionally, we confirm that the \textit{unpolarized} cross-sections exhibit minimal dependence on the spin-parity quantum numbers. Therefore, the discussions provided for Figs.~\ref{FIG2} and ~\ref{FIG3} remain applicable to other quantum-number cases without loss of generality.

\begin{figure}[t]
\begin{tabular}{cc}
\topinset{(a)}{\includegraphics[width= 8.5cm]{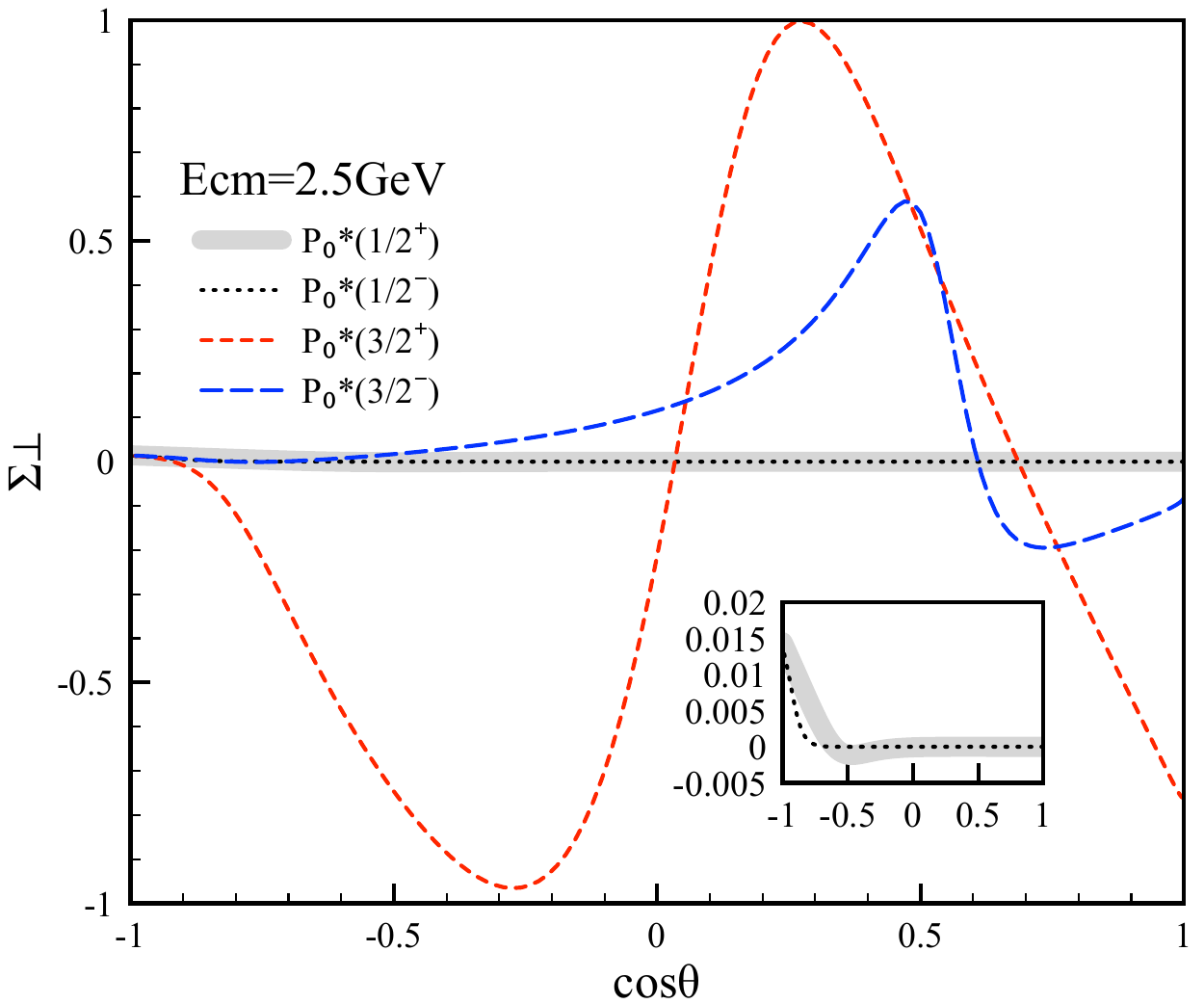}}{0.5cm}{-2cm}
\topinset{(b)}{\includegraphics[width= 8.5cm]{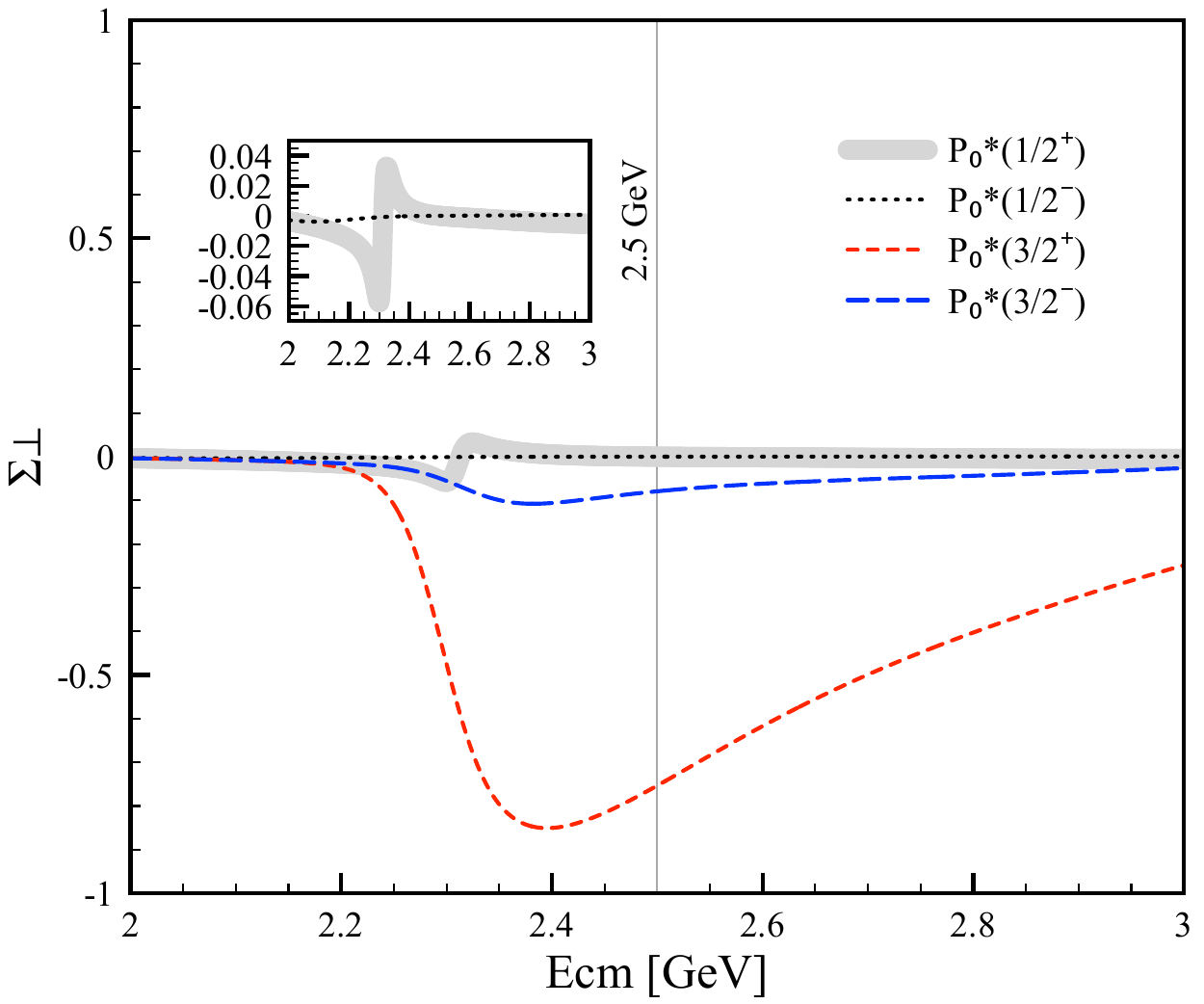}}{0.5cm}{-2cm}
\end{tabular}
\caption{(a) Recoil-proton spin asymmetry (RSA) in Eq.~(\ref{eq:ASYM}) as functions of $E_\mathrm{cm}$ for different spin-parity quantum numbers of $P^*0$, with $\bm{\epsilon}{K^*}$ perpendicular to the reaction plane. (b) $\Sigma_\perp$ as functions of $\cos\theta$ at $E_\mathrm{cm}=2.5$ GeV following the same approach.}       
\label{FIG4}
\end{figure}

Finally, we would like to propose a physical observable for determining the spin-parity quantum number of the $P^*_0$ resonance. The polarizations of the involved particles, such as the target and recoil nucleons, and $K^*$, are invaluable for this determination. However, achieving a definite neutron polarization inside the deuteron target poses challenges in experimental setups. As discussed earlier, to maximize the $S/N$, it's advantageous to set the $K^*$ polarization as $\bm{k}_K\cdot\bm{\epsilon}_{K^*}=0$ (perpendicular configuration). Consequently, we rely on controlling the recoil-proton polarization for quantum-number determination. Hence, we introduce the recoil-proton spin asymmetry (RSA) defined as follows, with $\bm{\epsilon}_{K^*}$ perpendicular to the reaction plane:
\begin{eqnarray}
\label{eq:ASYM}
\Sigma_\perp\equiv\frac{d\sigma_\uparrow/d\cos\theta-d\sigma_\downarrow/d\cos\theta}{d\sigma_\uparrow/d\cos\theta+d\sigma_\downarrow/d\cos\theta}\Big|_{\bm{k}_K\cdot\bm{\epsilon}_{K^*}=0},
\end{eqnarray}
where the notation ($\uparrow,\downarrow$) denotes the (up, down)-spin polarization of the recoil proton, respectively. In panel (a) of Fig.~\ref{FIG4}, we illustrate the RSAs as functions of $\cos\theta$ at $E_\mathrm{cm}=2.5$ GeV. Note that the curves mainly comprise contributions from the $s$-channel and $u$-channel, as the $t$-channel contribution is eliminated by fixing the polarization as $\bm{k}_K\cdot\bm{\epsilon}_{K^*}=0$. The spin-$1/2$ cases exhibit much smaller asymmetry compared to the spin-$3/2$ ones. The $P^*_0(1/2^\pm)$ amplitudes show weak angular dependence and are relatively small due to their Lorentz structures in the vicinity of $E_\mathrm{cm}=2.5$ GeV as follows:
\begin{eqnarray}
\label{eq:RES}
u^\dagger(p_2)\gamma_0\gamma_2\left(E_\mathrm{cm}\gamma_0\pm M_{P^*_0}\right)\gamma_5u(p_1)\sim A(E_\mathrm{cm}\pm M_{P^*_0})+B(E_\mathrm{cm}\mp M_{P^*_0})(\cos\theta,\sin\theta),
\,\,\,\,(A,B)\in R.
\end{eqnarray}
On the contrary, the $P^*_0(3/2^\pm)$ amplitudes exhibit highly nontrivial angular dependencies arising from the spin-sum of the Rarita-Schwinger field and the kinematic factor $\bm{k}_K\cdot\bm{k}_{K^*}\propto t$, as shown in Eq.~(\ref{eq:AMPS}). Consequently, the spin-parity quantum numbers can be determined from $\Sigma_\perp(\cos\theta)$ for the spin-$3/2$ cases, whereas fixing the parity for the spin-$1/2$ cases proves to be more challenging. In panel (b) of Fig.~\ref{FIG4}, the numerical results for the RSA as functions of $E_\mathrm{cm}$ are presented. Each spin-parity case exhibits a distinctive curve shape above $E_\mathrm{cm}\approx2.2$ GeV. Once more, the spin-$3/2$ cases demonstrate different energy dependence, whereas the spin-$1/2$ ones remain small and flat due to the same reasons as discussed above.

\section{Summary and future perspectives}
In our current study, we delve into the peak-like structure observed around $\sqrt{s}\sim2.5$ GeV in the reaction $K^+n\to K^{*0}p$. We consider the potential $S=+1$ resonance $P^{+*}_0\equiv P^*_0$ as an excited state within the extended vector-meson and baryon ($VB$) antidecuplet, particularly in the $s$ channel. To explore this, we employ the effective Lagrangian method in conjunction with the Regge approach at the tree-level Born approximation. We examine various hyperon ($Y$) exchanges in the $u$ channel and pion ($\pi$) ones in the $t$ channel. Below, we list important findings of the present studies:
\begin{itemize}
\item We find that the currently available experimental data around $E_\mathrm{cm}=2.5$ GeV cannot be adequately explained solely by the tree-level Born diagrams incorporating Regge contributions as the background (BKG). However, they are well accounted for by a resonance state characterized by $M_{P^*_0}=2.5$ GeV and $\Gamma_{P^*_0}=100$ MeV, alongside its spin-parity quantum numbers $J^P=1/2^\pm$ and $3/2^\pm$. To further elucidate this, we estimate the combined coupling constants for the resonance, denoted as $g_{P^*_0}\equiv g_{KN{P^*_0}}g_{K^*N{P^*_0}}$, by fitting the data, paving the way for future theoretical studies.
\item The angular dependence of the reaction process, represented by $d\sigma/d\cos\theta$, is primarily influenced by the contributions from the $t$-channel $\pi$ and $u$-channel $\Lambda(1116)$. These contributions result in significant enhancements of both forward and backward scattering in the center-of-mass (cm) frame, which gradually diminish as $E_\mathrm{cm}$ increases. However, in contrast, for all spin-parity states of $P^*_0$, it becomes evident that the resonance contribution is comparatively smaller than the background (BKG) and only becomes notable around $E\mathrm{cm}=2.5$ GeV, particularly in the vicinity of $\cos\theta=0$.
\item To accurately study the resonance contribution, we have devised an experimental approach aimed at enhancing the signal-to-noise ratio ($S/N$) by taking into account the structure of the predominant background signals (BKGs). By concentrating on the forward-scattering region and orienting the polarization of the $K^*$ perpendicular to the reaction plane, it becomes feasible to effectively circumvent the effects originating from the $u$ channel and deactivate those from the $t$ channel, respectively. This strategy yields a $S/N$ value close to unity. As the scattering angle increases towards $\cos\theta\approx0$, the contamination stemming from the $u$-channel becomes more pronounced, yet the $S/N$ value remains reasonably high.
\item Considering the ongoing reaction process, we delve into exploring a polarized observable known as the recoil-proton spin asymmetry (RSA) with $\bm{k}_K\cdot\bm{\epsilon}_{K^*}=0$ to discern the spin-parity quantum numbers of the $P^*_0$. We discern distinctive angular dependencies of the RSAs for the $J^P=3/2^\pm$ states. However, determining the parity for the spin-$1/2$ states proves to be challenging due to their weak and minimal angular dependencies at $E\mathrm{cm}=2.5$ GeV. Furthermore, we analyze the energy dependencies of the RSA for prospective experiments.
\end{itemize}

We wish to underscore that the potential $S=+1$ resonance state $P^*_0$, proposed to account for the currently available data on $K^+n\to K^{*0}p$, holds promise for illuminating our understanding of exotic phenomena in QCD, surpassing traditional chiral interaction theories. The conjecture that $P^*_0$ pertains to the resonance of the extended $VB$ antidecuplet, akin to the $PB$ antidecuplet for $\Theta^+$, remains inadequately elucidated within existing theories. Nevertheless, ongoing discourse contemplates plausible explanations for these antidecuplet resonances~\cite{NAMYANG}. Furthermore, more realistic theoretical studies will be conducive to future experimental analyses, i.e., $K^+n\to K^{*0}p\to\pi^{0,\mp} K^{0,\pm} p$ for instance. Such endeavors will yield valuable insights into $P^*_0$, particularly in the $K^0p$ invariant-mass domain. Relevant investigations are underway and will be detailed elsewhere.
\section*{Acknowledgment}
The authors extend their gratitude for the fruitful discussions with Jung Keun Ahn (Korea University), Shin Hyung Kim (Kyungpook National University), Ghil-Seok Yang (Hoseo University), Kanchan Pradeepkumar Khemchandani (Federal University of São Paulo), and Alberto Martinez Torres (University of São Paulo). This work received support from the National Research Foundation of Korea (NRF) grants funded by the Korean government (MSIT) (2018R1A5A1025563, 2022R1A2C1003964, and 2022K2A9A1A06091761).

\end{document}